\documentclass{article}
\usepackage{spconf,amsmath,graphicx}
\usepackage{multirow,subfigure, verbatim}

\def\x{{\mathbf x}}
\def\y{{\mathbf y}}
\def\v{{\mathbf v}}

\title{ADRN: Attention-based Deep Residual Network for Hyperspectral Image Denoising}
\name{Yongsen Zhao, Deming Zhai$^*$\thanks{Corresponding author. This work is supported by the National Science Foundation of China under Grants 61922027 and 61932022.}, Junjun Jiang, Xianming Liu}
\address{School of Computer Science and Technology, Harbin Institute of Technology, Harbin, China, 150001}
\begin{document}
%
\maketitle

\vspace{-0.3cm}
\begin{abstract}
\quad Hyperspectral image (HSI) denoising is of crucial importance for many subsequent applications, such as HSI classification and interpretation. In this paper, we propose an attention-based deep residual network to directly learn a mapping from noisy HSI to the clean one. To jointly utilize the spatial-spectral information, the current band and its $K$ adjacent bands are simultaneously exploited as the input. Then, we adopt convolution layer with different filter sizes to fuse the multi-scale feature, and use shortcut connection to incorporate the multi-level information for better noise removal. In addition, the channel attention mechanism is employed to make the network concentrate on the most relevant auxiliary information and features that are beneficial to the denoising process best. To ease the training procedure, we reconstruct the output through a residual mode rather than a straightforward prediction. Experimental results demonstrate that our proposed ADRN scheme outperforms the state-of-the-art methods both in quantitative and visual evaluations. 
\end{abstract}
\begin{keywords}
HSI denoising, Spatial-spectral, Channel attention, Residual learning
\end{keywords}
%

\vspace{-0.2cm}
\section{Introduction}
\label{sec:intro}

Hyperspectral image (HSI) data contains abundant saptial and spectral information, which makes it have a wide range of applications. Nevertheless, because of the senosr restriction and atmospheric interference, HSIs often suffer from various types of noise, such as Gaussian noise, stripe noise and dead lines, etc \cite{rasti2018noise}. Thus, it is essential to reduce the noise in HSIs in order to facilitate the following high-level analysis tasks.

The goal of HSI denoising is to recover a clean image \textbf{$\x$} from a noisy observation \textbf{$\y$}. The degradation model can be formulated as \textbf{$\y=\x+\v$}, where \textbf{$\v$} is additive white Gaussian noise (AWGN) with standard deviation $\sigma$ in general. To address this ill-posed inverse problem, the prior knowledge about $\x$ needs to be adopted to constrain the solution space. Over the past decades, in the literature, a variety of reasonable priors have been proposed for HSI denoising, such as total variation, non-local self-similarity, sparse representation, low-rank model and so on. For example, Maggioni \textit{et al.} \cite{BM4D} proposed an algorithm called BM4D which exploits the local correlation in each cube and the non-local correlation between different cubes. Considering the high spectral correlation across bands and high spatial similarity within each band, Renard \textit{et al.} \cite{LRTA} proposed a low-rank tensor approximation method (LRTA), which performs both spatial low-rank approximation and spectral dimensionality reduction. Besides, Zhang \textit{et al.} \cite{LRMR} proposed an efficient HSI restoration method based on low-rank matrix recovery (LRMR). Chang \textit{et al.} \cite{LLRT} claimed that the non-local self-similarity was the key ingredient for denoising, and proposed a unidirectional low-rank tensor recovery method to capture the intrinsic structure correlation in HSIs. To combine both the spatial non-local similarity and global spectral low-rank property, He \textit{et al.} \cite{NG-Meet} proposed a unified spatial-spectral paradigm for HSI denoising called NG-Meet. The major drawback of the above mentioned approaches is that they are time-consuming due to the complex optimization process, which prevents their usage in practice. In addition, these manually introduced prior knowledge only reflect the characteristics of a certain respect of the data, so the representation ability of these methods is limited.

\begin{figure*}[htb]
\centering
\includegraphics[width=0.95\linewidth]{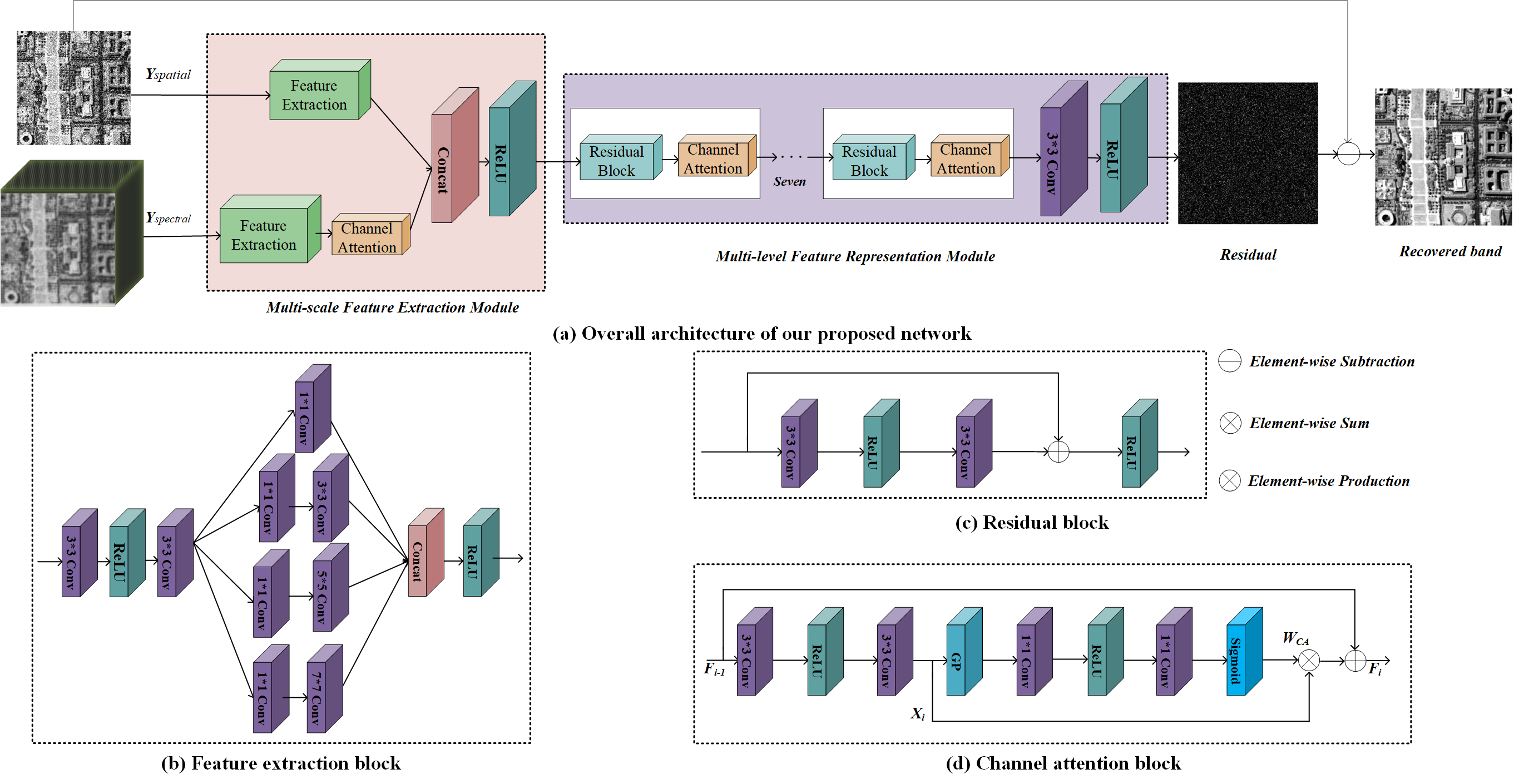}\
\caption{The detailed structure of our ADRN}
\vspace{-0.5cm}
\end{figure*}

Recently, deep learning based approaches have been proposed for hyperspectral image denoising. Yuan \textit{et al.} \cite{HSID-CNN} utilized both the spatial and spectral information to recover the clean image through multi-scale feature extraction and multi-level feature representation by neural networks. Zhang \textit{et al.} \cite{SSGN} proposed a spatial-spectral gradient network for mixed noise removal in HSIs, in consideration of the  spatial structure directionality and spectral differences. Although these methods achieve impressive denoising results, there is still much potential to explore and promote this domain forward.

A feasible strategy is to explore the most relevant part of the auxiliary spectral information to make full use of the spectral low-rank property, and make the network adaptively learn significant features.
In view of this point, in this paper, we introduce an attention-based deep residual convolutional neural network (ADRN) for HSI denoising. Both a single band and its $K$ adjacent bands are simultaneously fed into the network to take full advantage of the spatial-spectral information.
  Convolution layers with different sizes of reception field are adopted to extract multi-scale spatial and spectral feature respectively. Then, shortcut connections are built to enable the information flow from the fused feature representation to the final residual output, which can reduce the traditional degradation and feature vanish problem. More importantly, to increase the ability of discriminative learning, we integrate the channel attention mechanism into the network to make it more aware of the information that is more relevant and features that are more crucial. \textit{To the best of our knowledge, this is the first work in HSI denoising that considers the attention mechanism.} Compared with start-of-the-art methods, our proposed ADRN scheme achieves superior performance in both quantitative and visual evaluations.

\vspace{-0.3cm}
\section{Methodology}

\begin{table*}
\scriptsize
\centering
\caption{Quantitative performance comparison of the denoising results}
\begin{tabular}{ccccccccc}
\hline \hline
Noise Level          & Criterion & LRTA \cite{LRTA} & BM4D \cite{BM4D} & LRMR \cite{LRMR} & HSID-CNN \cite{HSID-CNN} & LLRT \cite{LLRT} & NG-Meet \cite{NG-Meet} & Proposed \\
\hline
\multirow{2.2}{*}{$\sigma_n=5$}   & MPSNR & 39.009$\pm$0.0034 & 41.188$\pm$0.0023 & 40.878$\pm$0.0036 & 41.684$\pm$0.0025 & 41.532$\pm$0.0054 &\textbf{41.781$\pm$0.0052} & \underline{41.580}$\pm$0.0043 \\ \cline{2-9}
                                & MSSIM & 0.9926$\pm$0.0002  & 0.9962$\pm$0.0001  & 0.9952$\pm$0.0001  &  0.9966$\pm$0.0001 & \underline{0.9968}$\pm$0.0001  & 0.9966$\pm$0.0001 & \textbf{0.9972$\pm$0.0001} \\ \hline
\multirow{2.2}{*}{$\sigma_n=25$}   & MPSNR &  30.672$\pm$0.0033 & 31.136$\pm$0.0025  & 33.029$\pm$0.0023  & 33.050$\pm$0.0028  & 34.701$\pm$0.0097  &\underline{35.366}$\pm$0.0094 & \textbf{35.527$\pm$0.0104} \\ \cline{2-9}
                                 & MSSIM & 0.9629$\pm$0.0002  & 0.9685$\pm$0.0002  & 0.9809$\pm$0.0001  & 0.9813$\pm$0.0001  & 0.9862$\pm$0.0 001  & \underline{0.9880}$\pm$0.0001 & \textbf{0.9902$\pm$0.0001} \\ \hline
\multirow{2.2}{*}{$\sigma_n=50$}   & MPSNR &  26.832$\pm$0.0052 &  26.752$\pm$0.0034 &  28.806$\pm$0.0043 & 28.968$\pm$0.0039  & 30.759$\pm$0.0115  & \underline{31.669}$\pm$0.0139 & \textbf{32.070$\pm$0.0102} \\ \cline{2-9}
                                 & MSSIM &  0.9246$\pm$0.0001 & 0.9208$\pm$0.0002  & 0.9532$\pm$0.0001  &  0.9536$\pm$0.0001 & 0.9705$\pm$0.0001  & \underline{0.9752}$\pm$0.0001 & \textbf{0.9796$\pm$0.0001} \\  \hline
\multirow{2.2}{*}{$\sigma_n=75$}   & MPSNR & 24.682$\pm$0.0054  & 24.261$\pm$0.0035  & 26.306$\pm$0.0046  & 26.753$\pm$0.0039  & 28.385$\pm$0.0134  & \underline{29.116}$\pm$0.0147 & \textbf{29.862$\pm$0.0175} \\ \cline{2-9}
                                 & MSSIM & 0.8866$\pm$0.0001  & 0.8670$\pm$0.0001  & 0.9192$\pm$0.0001  &  0.9273$\pm$0.0001 & 0.9525$\pm$0.0002  & \underline{0.9594}$\pm$0.0001 & \textbf{0.9673$\pm$0.0001} \\ \hline
\multirow{2.2}{*}{$\sigma_n=100$}   & MPSNR & 23.175$\pm$0.0048  &  22.577$\pm$0.0054 & 24.310$\pm$0.0047  & 25.296$\pm$0.0043  & 26.712$\pm$0.0145  & \underline{27.756}$\pm$0.0083 & \textbf{28.239$\pm$0.0176} \\ \cline{2-9}
                                 & MSSIM & 0.8494$\pm$0.0003  & 0.8119$\pm$0.0002  & 0.8799$\pm$0.0002  & 0.9014$\pm$0.0001  & 0.9328$\pm$0.0001  & \underline{0.9454}$\pm$0.0001 & \textbf{0.9535$\pm$0.0002} \\ \hline
\multirow{2.2}{*}{$\sigma_n=rand(25)$}   & MPSNR & 28.843$\pm$0.0025  & 34.424$\pm$0.0034  & 36.094$\pm$0.0033  & \underline{37.367}$\pm$0.0028  & 34.360$\pm$2.6908  & 36.040$\pm$0.3682 & \textbf{37.301$\pm$0.1633} \\ \cline{2-9}
                                       & MSSIM &  0.9331$\pm$0.0001 & 0.9833$\pm$0.0002  & 0.9856$\pm$0.0001  & \underline{0.9916}$\pm$0.0001  & 0.9718$\pm$0.0275  & 0.9904$\pm$0.0001 & \textbf{0.9917$\pm$0.0004} \\ \hline
\multirow{2.2}{*}{$\sigma_n=Gau(200, 30)$}& MPSNR & 28.200$\pm$0.0023 & 34.109$\pm$0.0037 & 35.962$\pm$0.0025 & \underline{36.804}$\pm$0.0029 & 28.635$\pm$0.0019 & 35.402$\pm$0.0053 & \textbf{37.722$\pm$0.0080} \\ \cline{2-9}
                                         & MSSIM &  0.9119$\pm$0.0002 & 0.9794$\pm$0.0001  &0.9893$\pm$0.0001   & \underline{0.9895}$\pm$0.0001  & 0.9094$\pm$0.000  & 0.9894$\pm$0.0001 &\textbf{0.9929$\pm$0.0001}  \\ \hline
\hline
\label{table:result}
\vspace{-0.9cm}
\end{tabular}
\end{table*}

In this section, we introduce in detail the proposed attention-based deep residual network for HSI denoising.
The overall architecture of our network is illustrated in Fig. 1(a). $Y_{spatial}$ represents an input noisy band and $Y_{spectral}$ denotes its $K$ adjacent bands. The multi-scale feature extraction module is in charge of acquiring the spatial contextual and spectral correlation information for further processing. Then the multi-level feature representation module is tailored to construct the residual noise. Finally,  the clean signal is obtained through subtracting the residual from the spatial input. In the following, we will elaborate blocks and loss function of our network.

\vspace{-0.2cm}
\subsection{Feature Extraction Block}
The ground objects in HSIs have various sizes in different regions naturally. This fact implies that our denoising network should be able to capture the contextual information of multiple scales. Inspired by Inception \cite{Inception}, in our network, four types of convolution layers with reception field sizes--1, 3, 5, 7--are adopted to extract both the spatial and spectral features, as described in Fig. 1(b). Furthermore, to avoid the expensive computation burden and accelerate the speed in test, a $1\times 1$ convolution layer is inserted to reduce the channel dimension when the filter size is more than 1. 


\vspace{-0.2cm}
\subsection{Residual Block}
As the network goes deeper, information extracted from the early stage of the network may vanish or ”wash out” by  the time it reaches the output layer \cite{dense}. In addition, the deeper networks often suffer from gradient vanishing problem, which makes the training process slow or even divergent. To address these problems, we adopt the shortcut connection from ResNet \cite{ResNet} to directly pass the early feature map to the later layers, as illustrated in Fig. 1(c). This greatly increases the flow of information and thus contributes to the prediction of residual noise and the back propagation of gradients, thereby accelerating the training process.

\vspace{-0.2cm}
\subsection{Channel Attention Block}
The traditional CNN treats each channel of a feature map equally, which lacks discriminativa learning ability across channels and thus inhibits the representation power of deep networks.  We observe that feature maps extracted from  the spectral input contribute differently to the final denoising result, and some of them may be not that beneficial. Thus, what our network learns should concentrate on the significant features. Moreover, in our residual learning strategy, convolution kernels that are responsible for high-frequency extraction should be paid more attention to facilitate the prediction of noise. In view of these concerns, we introduce a channel attention block to adaptively modulate feature representation.

The structure of our channel attention block (CAB) is illustrated in Fig. 1(d). For the $i$-th CAB, we have
\begin{equation}
\small
    F_i = F_{i-1} + W_{CA} * X_i
\end{equation}
where $F_i$ and $F_{i-1}$ are the input and output feature map respectively, $X_i$ is the residual component acquired by two stacked convolution layer equipped with filter size of $3\times 3$:
\begin{equation}
\small
    X_i = W_2*\delta(W_1*F_{i-1})
\end{equation}
where $W_1$ and $W_2$ are weight sets and $\delta$ denotes the ReLU function.
$W_{CA}$ is the learned calibration weight, for which we exploit the global average pooling on $X_{i}$ first. A $1\times 1$ convolution layer with ReLU is followed to downsample the channel number by the ratio $r$. Then, the channel number is increased back to the original amount through a $1\times 1$ convolution layer with Sigmoid to guarantee $W_{CA}$ lies in [0,1]:
\begin{equation}
\small
    W_{CA} = Sigmoid(W_4*\delta(W_3*GP(X_i)))
\end{equation}
where $W_3$ and $W_4$ are weight sets and $GP$ means the global average pooling operation.

\vspace{-0.2cm}
\subsection{Residual Learning and Loss Function}

In order to avoid the degradation phenomenon as the network goes deeper and ensure the stability and effectiveness of the training process, our network does not directly predict the clean image, but outputs a residual noise $R$:
\begin{equation}
\small
    R = F(\Theta, Y_{spatial}, Y_{spectral})
\end{equation}
where $\Theta$ denotes model parameters learned by back propagation algorithm.
Then the restored clean image $\hat X$ can be obtained by subtracting residual noise from the spatial input:
\begin{equation}
\small
   \hat X = Y_{spatial} - R 
\end{equation}

The loss function of our training process consists of two parts: reconstruction loss $L_{rec}$ and regularization loss $L_{reg}$:
\begin{equation}
\small
    L_{total} = \lambda L_{rec} + L_{reg}
\end{equation}
where $\lambda$ controls the trade-off between two terms.
$L_{rec}$ aims to ensure the restored result approximate to the ground truth:
\begin{equation}
\small
    L_{rec} = \frac{1}{NHW}\sum_{i=1}^N||\hat X^i - X^i||_2^2
\end{equation}
while $L_{reg}$ is used to enforce the residual noise satisfy a zero-mean distribution.:
\begin{equation}
\small
    L_{reg} = (\frac{1}{NHW}\sum_{i=1}^N\sum_{h=1}^H\sum_{w=1}^W R_{hw}^i)^2
\end{equation}
where $N$ denotes the number of training batch, $H$ and $W$ mean the height and width of training images.

\vspace{-0.2cm}
\subsection{Implementation Details}
The adjacent band number $K$ is set to 64, the downsample ratio $r$ is set to 10 as in \cite{RCAN} and the trade-off parameter $\lambda$ is equal to 10 during all the training procedure. We use the truncated normal distribution to initialize the weights and train the network from scratch. In optimization, we exploit Adam \cite{adam} with a mini-batch size of 382 (two times of the band number), while the parameters for Adam are set as $\beta_1 = 0.9$, $\beta_2 = 0.999$ and $\epsilon = 1e-8$, which follow the default setting in TensorFlow \cite{TensorFlow}. The learning rate starts from 0.0001 and decays exponentially every certain training steps (such as 5000). The total iteration is roughly about 300,000 times.
\section{Experiments}

\begin{figure*}[htb]
\centering
\includegraphics[width=1\linewidth]{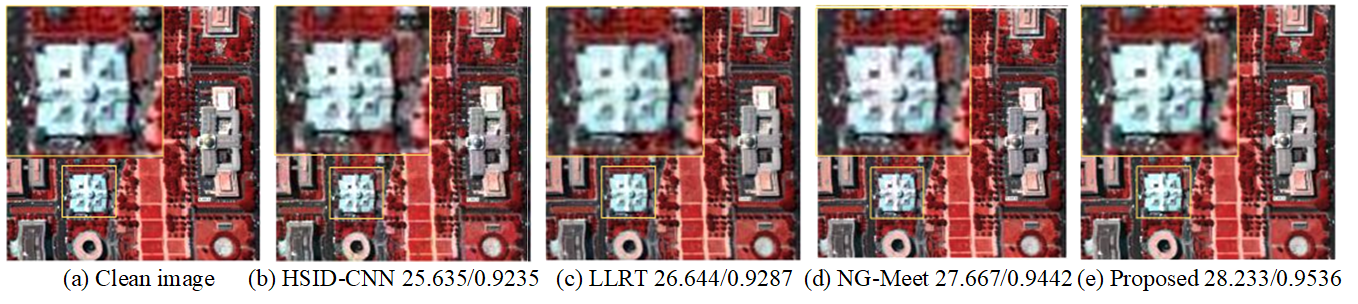}
\vspace{-0.8cm}
\caption{Quantitative and visual comparison for Washington DC Mall image with $\sigma_n=100$}
\vspace{-0.2cm}
\end{figure*}

\begin{figure*}[htb]
\centering
\includegraphics[width=1\linewidth]{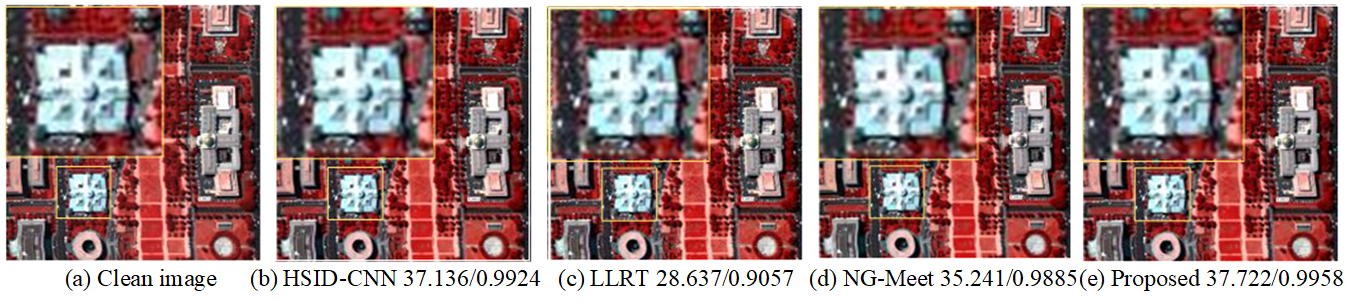}
\vspace{-0.4cm}
\caption{Quantitative and visual comparison for Washington DC Mall image with $\sigma_n=Gau(200, 30)$}
\vspace{-0.4cm}
\end{figure*}

In this section, extensive experimental results are provided to validate the effectiveness of our method. Several state-of-the-art methods are used for comparison, including: BM4D \cite{BM4D}, low-rank tensor approximation (LRTA) \cite{LRTA},  LRMR \cite{LRMR}, HSID-CNN \cite{HSID-CNN}, LLRT \cite{LLRT} and NG-Meet \cite{NG-Meet}. MPSNR \cite{MPSNR} and MSSIM \cite{MSSIM} are served as the evaluation criterion. Better HSI denoising results lead to higher MPSNR and MSSIM.

We follow exactly the same setting in deep model training and test as HSID-CNN \cite{HSID-CNN}. We use the Washington DC Mall image with a size of $1280\times 303\times 191$ to train our model, out of which we select $200\times 200\times191$ for testing and the other part of $1080\times 303\times 191$ for training. First, we utilize the ENVI software to normalize the gray values of each HSI band to [0,1]. Then we crop $20\times 20$ patches from the training part with the stride of 5. The simulated noisy patches are generated through imposing additive white Gaussian noise (AWGN) with standard deviation of [5, 25, 50, 75, 100] to formulate the training data. 
For the simulated HSI denoising process, three types of noise are employed: First, different bands have the same noise intensity. For instance, $\sigma_n$ is set from 5 to 100, as listed in Table \ref{table:result}. Second, the noise intensity of different bands conforms a random probability distribution, labeled as \textit{rand(25)}. Third, for different bands, the noise intensity is also different but varies like a Gaussian distribution centered at the middle band:
\begin{equation}
\small
   \sigma_n = \beta \sqrt{\frac{exp\{-(k-B/2)^2/2\eta^2\}}{\sum_{k=1}^{B}exp\{-(k-B/2)^2/2\eta^2\}}} 
\end{equation}
where $\beta=200$, $\eta=30$ and $B=191$ in our settings.

The averages and standard deviations of MPSNR and MSSIM are obtained by repeating 10 runs of compared methods. The best performance for each quality criterion is marked in bold and the second-best one is underlined. Compared with other algorithms, the proposed ADRN achieves the highest MPSNR and MSSIM values in almost all noisy levels except the case $\sigma_n=5$. Under such a small noise level, all methods achieve a relatively good performance and the gap is small. In contrast, as the noise level goes higher and more complicated, our approach clearly outperforms other algorithms.

It is worth noting that NG-Meet achieves the best HSI denoising performance in the literature. However, it assumes that noise follows independently and identically distributed (i.i.d) Gaussian distribution, and its performance dropped dramatically when encountering non-i.i.d. noise. 

To further demonstrate the effectiveness of our proposed method, Fig. 2 and Fig. 3 show the pseudo-color images of the test data (composed of bands 57, 27 and 17) after denoising in the case $\sigma_n=100$ and $\sigma_n=Gau(200, 30)$ respectively. The MPSNR and MSSIM values of each method are marked under the denoised images. Although LLRT and NG-Meet show a good noise reduction ability under the uniform noise intensities, it does not work well under unequal noise intensities for different bands. Our proposed method achieves the best performance in objective and subjective evaluations, which demonstrate the effectiveness of our proposed method.


\vspace{-0.2cm}
\section{Conclusion}
In this paper, we presented an attention-based deep residual network for HSI denoising. Both the spatial information and its adjacent bands are simultaneously assigned to the model to fully exploit the spatial-spectral structural correlation. Then, through incorporating the convolution layer of various reception fields, shortcut connection, and channel attention mechanism, we formulate a multi-scale feature extraction module and a multi-level feature representation module to respectively capture both the multi-scale spatial-spectral feature and fuse feature representations with different levels for the final restoration. Furthermore, we adopt the residual learning strategy to ensure the stability and efficiency of the training procedure. The simulated experiment indicated that our propose ADRN outperforms  mainstream methods in both quantitative and visual evaluations.



\bibliographystyle{IEEEtran}
\bibliography{refs}

\end{document}